\documentclass[letter,traditabstract]{aa} 
\usepackage{graphicx}
\usepackage{txfonts}
\newcommand{\tablenotea}[1]{\parbox{8.8cm}{ \indent
\footnotesize{\textsc{Note.--}~#1}}}
\newcommand{\jms}{J.~Mol.~Spectr.}                 %
\newcommand{\cpl}{Chem.~Phys.~Lett.}               %
\newcommand{\molphys}{Mol.~Phys.}                  %
\newcommand{\apl}{Astrophys.~Lett.}                %

\begin{document}
\title{HIFI\thanks{Herschel is an ESA space
observatory with science instruments provided by European-led
Principal Investigator consortia and with important participation
from NASA.} detection of HF in the carbon star envelope IRC
+10216}
\titlerunning{Hydrogen halides in IRC +10216}
\authorrunning{Ag\'undez et al.}
\author{M. Ag\'undez\inst{1}, J. Cernicharo\inst{2}, L. B. F. M. Waters\inst{3,4}, L. Decin\inst{4,5}, P. Encrenaz\inst{6}, D. Neufeld\inst{7}, D. Teyssier\inst{8}, and F. Daniel\inst{2}}
\institute{LUTH, Observatoire de Paris-Meudon, 5 Place Jules
Janssen, 92190 Meudon, France; \email{marcelino.agundez@obspm.fr}
\and Departamento de Astrof\'isica, Centro de Astrobiolog\'ia,
CSIC-INTA, Ctra. de Torrej\'on a Ajalvir km 4, 28850 Madrid, Spain
\and SRON Netherlands Institute for Space Research, Sorbonnelaan
2, 3584 CA Utrecht, The Netherlands \and Sterrenkundig Instituut
Anton Pannekoek, University of Amsterdam, Science Park 904,
NL-1098, Amsterdam, The Netherlands \and Instituut voor
Sterrenkunde, Katholieke Universiteit Leuven, Celestijnenlaan
200D, 3001 Leuven, Belgium \and LERMA, Observatoire de Paris, 61
Av. de l'Observatoire, 75014 Paris, France \and Department of
Physics and Astronomy, The Johns Hopkins University, 3400 North
Charles Street, Baltimore, MD 21218, USA \and European Space
Astronomy Centre, Urb. Villafranca del Castillo, PO BOX 50727,
28080, Madrid, Spain}

\date{Received; accepted}


\abstract
{We report the detection of emission in the $J=1-0$ rotational
transition of hydrogen fluoride (HF), together with observations
of the $J=1-0$ to $J=3-2$ rotational lines of H$^{35}$Cl and
H$^{37}$Cl, towards the envelope of the carbon star IRC +10216.
High-sensitivity, high-spectral resolution observations have been
carried out with the HIFI instrument on board \emph{Herschel},
allowing us to resolve the line profiles and providing insights
into the spatial distribution of the emission. Our interpretation
of the observations, with the use of radiative transfer
calculations, indicates that both HF and HCl are formed in the
inner regions of the envelope close to the AGB star.
Thermochemical equilibrium calculations predict HF and HCl to be
the major reservoirs of fluorine and chlorine in the atmospheres
of AGB stars. The abundances relative to H$_2$ derived for HF and
HCl, $8 \times 10^{-9}$ and $10^{-7}$ respectively, are
substantially lower than those predicted by thermochemical
equilibrium, indicating that F and Cl are likely affected by
significant depletion onto dust grains, although some chlorine may
be in the form of atomic Cl. The H$^{35}$Cl/H$^{37}$Cl abundance
ratio is 3.3 $\pm$ 0.3. The low abundance derived for HF in IRC
+10216 makes it likely that the fluorine abundance is not enhanced
over the solar value by nucleosynthesis in the AGB star, although
this conclusion may not be robust because the HF abundance we
derive is a lower limit to the elemental abundance of F. These
observations suggest that both HF and HCl should be detectable
through low $J$ rotational transitions in other evolved stars.}
{}
{}
{}
{}

\keywords{astrochemistry --- line: identification --- molecular
processes --- stars: AGB and post-AGB --- circumstellar matter ---
stars: individual (IRC +10216)}

\maketitle
%

\section{Introduction}

Light hydrides such as HF and HCl are very difficult to study in
space by means of their rotational spectra, which lie in the
submillimeter and far-infrared spectral regions that are hard if
not impossible to observe from the ground owing to the opacity of
the terrestrial atmosphere. Moreover, the prospects for detection
are limited by the low cosmic abundance of fluorine and chlorine,
although HF and HCl are expected to be the major reservoirs of
each of these elements in many astrophysical environments.

Since it is one of the lightest molecules, HF has a rotational
spectrum that can only be observed from space. It was first
detected with the \emph{Infrared Space Observatory} in Sagittarius
B2 (\cite{neu97} 1997). Observations of HCl have been conducted in
molecular clouds (\cite{bla85} 1985; \cite{zmu95} 1995;
\cite{sch95} 1995; \cite{sal96} 1996) through its fundamental
rotational transition, lying at 625.9 GHz and still accessible
from ground, and in the diffuse cloud $\zeta$ Ophiuchi
(\cite{fed95} 1995) through an electronic transition at
ultraviolet wavelengths using the \emph{Hubble Space Telescope}.
That HF and HCl are very stable molecules makes them abundant in
the atmospheres of cool stars and in sunspots, where they have
long been observed through ro-vibrational transitions in the
infrared region (\cite{hal69} 1969; \cite{hal72} 1972;
\cite{rid84} 1984; \cite{jor92} 1992).

With the launch of the \emph{Herschel Space Observatory}, it has
been possible to perform very sensitive observations of the pure
rotational lines of HF and HCl in different types of interstellar
clouds (see \cite{neu10} 2010; \cite{phi10} 2010; \cite{son10}
2010; \cite{cer10a} 2010a; \cite{lis10} 2010). In this Letter, we
present sensitive observations of the $J=1-0$ rotational
transition of HF and of the $J=1-0$ to $J=3-2$ transitions of
H$^{35}$Cl and H$^{37}$Cl towards the carbon star envelope IRC
+10216. This is the first time that HF and three lines of the two
major isotopologues of HCl are observed at high spectral
resolution in IRC +10216\footnote{Several transitions of HCl were
observed at low spectral resolution with the SPIRE and PACS
instruments (\cite{cer10b} 2010b), and the $J$ =1-0 line was
observed with the CSO telescope (\cite{pen10} 2010).} allowing us
to resolve the line profiles and gain insight into the spatial
distribution of the molecules.

\section{Observations}

\begin{figure}
\begin{center}
\includegraphics[angle=-90,scale=.37]{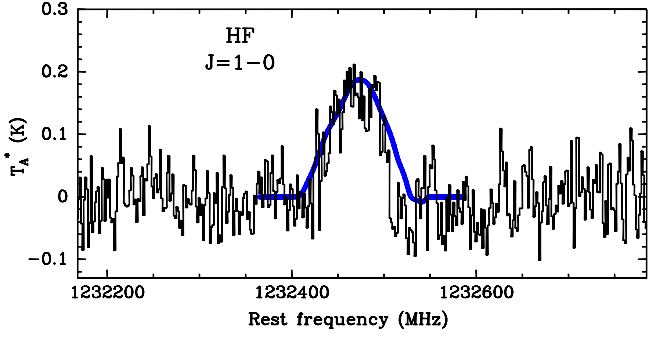}
\caption{HIFI spectrum of IRC +10216 showing the $J=1-0$
rotational line of HF (black histogram; spectral resolution is 1.5
MHz) and the line profile resulting from the radiative transfer
model (blue line).} \label{fig-hf-line} \vspace{-0.6cm}
\end{center}
\end{figure}

Our HIFI observations of the $J=1-0$ line of HF and of the $J=1-0$
to $J=3-2$ lines of H$^{35}$Cl and H$^{37}$Cl were obtained in May
and from October to December 2010, in the context of GT1
programmes to perform a line survey and to search for light
hydrides in IRC +10216. Data were taken in double beam-switching
mode with a spectral resolution of 1.1 MHz, and processed using
the standard Herschel pipeline up to level 2, which provides fully
calibrated spectra. The local oscillator was shifted in frequency
to identify any emission from the image band. Total integration
times were 43 minutes for the HF line, and ranged from 10 to 21
minutes for the HCl lines. The final spectra were smoothed to a
spectral resolution of 1.5 MHz and have antenna temperature rms
noise levels in the range 0.04--0.1 K. For details about the data
reduction, we refer to \cite{cer10c} (2010c).

\begin{table}[b]
\caption{HF and HCl observed line parameters in IRC +10216}
\label{table-lineparameters} \centering
\begin{tabular}{c@{\hspace{0.2cm}}c@{\hspace{0.3cm}}c@{}c@{}c}
\hline \hline
\multicolumn{1}{c}{}           & \multicolumn{1}{c}{$\nu_{\rm cal}$$^a$} & \multicolumn{1}{c}{$\nu_{\rm obs}$} & \multicolumn{1}{c}{v$_{\rm exp}$$^b$} & \multicolumn{1}{c}{$\int$$T_A^*$$d$v} \\
\multicolumn{1}{c}{Transition} & \multicolumn{1}{c}{(MHz)}               & \multicolumn{1}{c}{(MHz)}           & \multicolumn{1}{c}{(km s$^{-1}$)}     & \multicolumn{1}{c}{(K km s$^{-1}$)} \\
\hline
\multicolumn{5}{c}{HF} \\
$J=1-0$                           & 1232476.3 & 1232467.4(80) & 10.7(20) & 2.4(2) \\
\hline
\multicolumn{5}{c}{H$^{35}$Cl} \\
$J=1-0$$^c$                       &  625915.2 & --            & --       & 4.2(1) \\
$J=2-1$                           & 1251450.6 & 1251445.3(60) & 13.3(25) & 4.6(6) \\
$J=3-2$                           & 1876226.6 & 1876216.3(80) & 12.4(20) & 10.5(6) \\
\hline
\multicolumn{5}{c}{H$^{37}$Cl} \\
$J=1-0$$^c$                       &  624975.1 & --            & --       & 1.5(1) \\
$J=2-1$                           & 1249571.4 & 1249563.3(80) & 15.7(30) & 3.1(5) \\
$J=3-2$                           & 1873410.7 & 1873402.0(80) & 12.5(15) & 5.4(6) \\
\hline
\end{tabular}
\tablenotea{$^a$ Calculated frequencies of HF and HCl from
\cite{nol87} (1987) and \cite{caz04} (2004), respectively. $^b$
v$_{\rm exp}$ is the half width at zero intensity. Numbers in
parentheses are 1$\sigma$ uncertainties in units of the last
digits. $^c$ Line fitted to three components, with the center
frequencies fixed to the calculated values of the three hyperfine
components and the v$_{\rm exp}$ parameter fixed at 14.5 km
s$^{-1}$.}
\end{table}

\begin{figure}
\begin{center}
\includegraphics[angle=0,scale=.46]{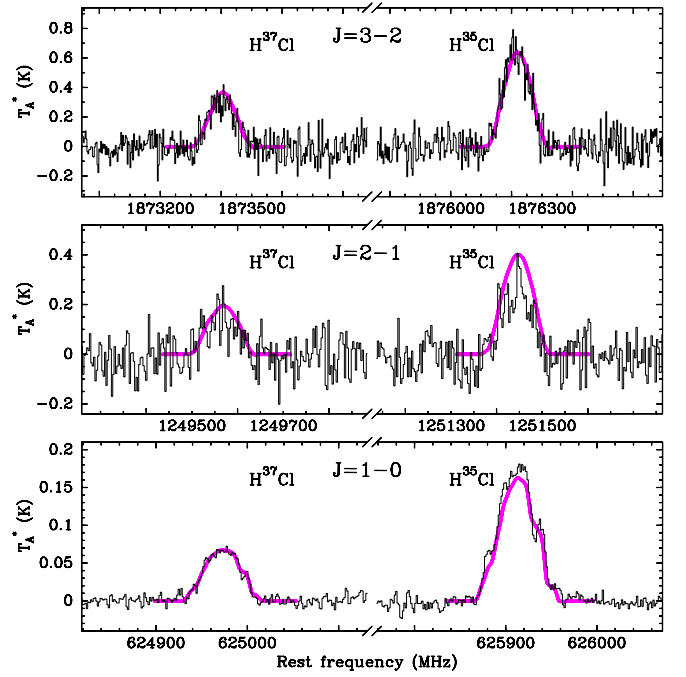}
\caption{HIFI spectra of IRC +10216 showing the first three
rotational lines of H$^{35}$Cl and H$^{37}$Cl (black histograms;
spectral resolution is 1.5 MHz for the $J=1-0$ and 3 MHz for the
$J=2-1$ and $J=3-2$ lines). Line profiles resulting from the
radiative transfer model are shown in magenta. For the $J=1-0$
line, the calculated intensity has been divided among the three
hyperfine components according to their line strengths.}
\vspace{-0.6cm} \label{fig-hcl-lines}
\end{center}
\end{figure}

\section{Results}

The $J=1-0$ line of HF and the $J=1-0$ to $J=3-2$ lines of
H$^{35}$Cl and H$^{37}$Cl are all clearly detected, free of
contamination by other lines (see Fig.~\ref{fig-hf-line} and
Fig.~\ref{fig-hcl-lines}). The differences between the observed
and calculated frequencies are not negligible, up to $\sim$10 MHz
(see Table~\ref{table-lineparameters}). They are, however, within
the errors and, expressed in Doppler equivalent velocity, have
values in the range 1.3--1.9 km s$^{-1}$, thus are much smaller
than the observed line widths (21--31 km s$^{-1}$). The detection
of HF, H$^{35}$Cl, and H$^{37}$Cl in IRC +10216 is therefore quite
secure.

The HCl hyperfine structure due to the nuclear quadrupole of the
$^{35}$Cl and $^{37}$Cl nuclei can only be barely distinguished in
the $J=1-0$ lines, but not in the higher $J$ lines owing to the
collapse of the hyperfine structure and to the line widening,
expressed in frequency, with increasing $J$. The $J=1-0$ line of
both isotopologues is composed of three hyperfine components but a
good determination of the parameters of each component was not
possible, so that we only give the total velocity integrated
intensity in Table~\ref{table-lineparameters}. The remaining lines
of HCl and that of HF have roughly parabolic shapes, which
indicates optically thick emission unresolved by the beam of HIFI.
The $J=1-0$ line of HCl was observed in IRC +10216 by \cite{pen10}
(2010) with the CSO. The smaller beam size of the CSO (13.5''), as
compared with that of HIFI (33.9''), implies a line intensity
enhancement of a factor of (33.9/13.5)$^2$ $\sim$6 for a
point-like source. This is essentially the value observed for the
$J=1-0$ line of H$^{35}$Cl after correcting for the beam
efficiency of each telescope (0.5 for the CSO and 0.75 for HIFI at
625.9 GHz). Therefore, the emission size of the HCl $J=1-0$ line
must be substantially smaller than the 13.5'' beam of the CSO.
This is consistent with the source size expected for the HCl
$J=1-0$ line based on its critical density ($\gtrsim$10$^7$
cm$^{-3}$), which is only exceeded in regions inside 10$^{15}$ cm
(0.5'' at a distance of 130 pc).

\section{Radiative transfer modelling}

To learn more about the excitation conditions and abundance of HF
and HCl in IRC +10216, we performed radiative transfer
calculations based on a multi-shell large velocity gradient (LVG)
formalism. The physical model consists of a spherical envelope of
gas and dust around an AGB star with a radius of 4 $\times$
10$^{13}$ cm and an effective temperature of 2330 K. The adopted
mass loss rate and distance to IRC +10216 are 2 $\times$ 10$^{-5}$
M$\odot$ yr$^{-1}$ and 130 pc, respectively. Further details about
the physical model will be given in a forthcoming paper (Ag\'undez
et al. in preparation).

To model the emission of the HF and HCl lines, we included the
first 10 and 15 rotational states within the ground and
first-excited vibrational states. The $v=1-0$ band lies at a
wavelength of 2.5 $\mu$m for HF and 3.5 $\mu$m for HCl. Level
energies were computed from the laboratory spectroscopic constants
of HF (\cite{oda99} 1999; \cite{ram96} 1996) and HCl (\cite{caz04}
2004; \cite{den97} 1997; \cite{leb94} 1994), neglecting the
hyperfine structure for this latter species. Dipole moments used
to compute the line intensities of HF are 1.826526 and 1.87368
Debye for pure rotational transitions within the $v=0$ and $v=1$
states, respectively (\cite{mue70} 1970; \cite{pie98} 1998), and
0.099735 Debye for the fundamental ro-vibrational transition
(\cite{pin85} 1985). For HCl, we used 1.109 and 1.139 Debye for
the $v=0$ and $v=1$ states, respectively (\cite{del71} 1971;
\cite{kai70} 1970), while for the $v=1-0$ vibrational band we
adopted 0.072961 Debye for H$^{35}$Cl and 0.073049 Debye for
H$^{37}$Cl (\cite{pin85} 1985). State-to-state rate constants for
rotational de-excitation of HF through inelastic collisions with
para and ortho H$_2$ were taken from \cite{gui11} (2011), whose
calculations cover the first seven rotational levels of HF and
extend up to a temperature of 200 K. An ortho-to-para ratio of 3
was adopted for H$_2$. For HF-He collisions, we adopted the values
calculated by \cite{ree05} (2005), covering the first ten levels
of HF up to 300 K. The use of collision rate constants with H$_2$,
which are larger than with He by up to one order of magnitude, is
critical for the excitation analysis of HF. We did not extrapolate
the rate constants in temperature, based on the small variation in
the collision cross sections for temperatures above 200-300 K (see
\cite{gui08} 2008, 2011). As rate constants for collisions of HCl
and H$_2$, we adopted the values calculated by \cite{neu94} (1994)
up to 300 K with He as collider, scaled up by a factor of 1.35 and
without extrapolating in temperature. Collision rates of HCl with
H$_2$ may differ from those with He, although it is difficult to
quantify these differences.

The computed line profiles are plotted in Fig.~\ref{fig-hf-line}
for HF and in Fig.~\ref{fig-hcl-lines} for H$^{35}$Cl and
H$^{37}$Cl. The agreement between observed and calculated line
profiles is very good, except for the $J=2-1$ transition of
H$^{35}$Cl and of H$^{37}$Cl, which are predicted to be somewhat
stronger than observed. The modest signal-to-noise ratio of these
lines and the lack of collision rate constants for HCl and H$_2$
could explain the discrepancies.

\begin{figure}
\begin{center}
\includegraphics[angle=-90,scale=.35]{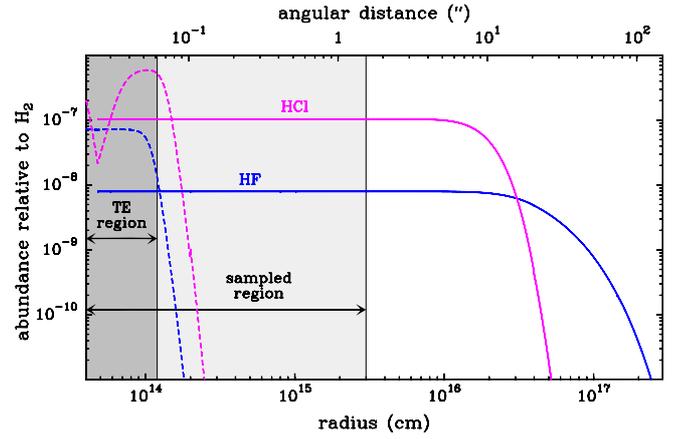}
\caption{Abundances relative to H$_2$ of HF and HCl (including
H$^{35}$Cl and H$^{37}$Cl) as a function of radius in IRC +10216.
Abundances plotted as solid lines have a constant value in the
inner envelope, derived from a fit to the observed lines (see
Fig.~\ref{fig-hf-line} and Fig.~\ref{fig-hcl-lines}), and decrease
in the outer envelope due to photodissociation, as computed with a
photochemical model. The region of the envelope sampled by the
observed lines of HF and HCl (see text) extends up to $3 \times
10^{15}$ cm (light shaded region). Dashed lines correspond to
abundances computed by thermochemical equilibrium, which only
holds up to $\sim$3 R$_*$ (dark shaded region).}
\label{fig-chemistry} \vspace{-0.6cm}
\end{center}
\end{figure}

We derive an abundance relative to H$_2$ in the inner regions of
the envelope of 8 $\times$ 10$^{-9}$ for HF and 8 $\times$
10$^{-8}$ for H$^{35}$Cl, with a H$^{35}$Cl/H$^{37}$Cl abundance
ratio of 3.3 $\pm$ 0.3, consistent with that derived from previous
observations of NaCl, KCl, and AlCl in IRC +10216 (\cite{cer87}
1987; \cite{cer00} 2000). The abundances of HF and HCl (H$^{35}$Cl
+ H$^{37}$Cl) are plotted in Fig.~\ref{fig-chemistry} as a
function of radius, for a model in which the initial abundances is
set by the observations, but its radial dependence is a model
prediction. Both HF and HCl are predicted to show a decrease in
abundance in the outer layers, owing to photodissociation by
interstellar ultraviolet photons. However, most of the
contribution to the observed line intensities ($\sim$80 \%) comes
from circumstellar regions inside $\sim$2 $\times$ 10$^{15}$ cm
for HF, and 2--4 $\times$ 10$^{15}$ cm for HCl, such that only the
abundances in regions within these radii are properly sampled by
the observed lines (see light shaded region in
Fig.~\ref{fig-chemistry}). The excitation of the levels involved
in the observed HF and HCl lines is dominated by inelastic
collisions with H$_2$, with the lines being subthermally excited
beyond $\sim$2 $\times$ 10$^{14}$ cm where the gas density is
below 10$^9$ cm$^{-3}$. Infrared pumping via the $v=1$ vibrational
state plays a non-negligible but minor role, enhancing the line
intensities by 5 \% (HF) and 15 \% (HCl) with respect to the case
where infrared pumping is neglected. Rotational lines within the
$v=1$ state are predicted to be too weak to be detectable with
HIFI, with predicted antenna temperatures of only $\sim$0.025 K
for the $v=1$ $J=3-2$ transition of H$^{35}$Cl and less than 0.005
K for all other such transitions. The abundances derived here are
estimated to be uncertain by a factor of two for both HF and HCl.
The error of the observations being 10-20~\%, 2-15~\% due to the
noise in the spectra (see Table~\ref{table-lineparameters}) and
10~\% due to the calibration of HIFI, most of the uncertainty
comes from the model. For HCl, the error could be significantly
larger if the collision rate constants with H$_2$ as collider are
substantially different from those with He. An increase in the
collision rate constants would lower the abundance derived for
HCl.

\section{Discussion}

The presence of both HF and HCl in the inner circumstellar regions
of IRC +10216 is consistent with the thermochemical equilibrium
(TE) calculations of the atmospheres of cool stars. In
Fig.~\ref{fig-chemistry}, we compare the abundances derived for HF
and HCl with those computed under the assumption of TE, adopting
solar elemental abundances (\cite{asp09} 2009). The assumption of
TE is only valid from the stellar photosphere out to a radius of
$\sim$3 R$_*$ (dark shaded region in Fig.~\ref{fig-chemistry}),
beyond which the decrease in density and temperature causes the
chemical reaction rates to drop rapidly and the molecular
abundances to freeze out to the values of the TE region
(\cite{agu06} 2006).

Hydrogen fluoride and AlF (whose abundance relative to H$_2$ is
$7.5 \times 10^{-9}$; \cite{agu09} 2009) are observed to be the
main fluorine-bearing species in the inner envelope of IRC +10216.
Thermochemical equilibrium calculations predict that HF is the
major reservoir of fluorine in the TE region around both oxygen-
and carbon-rich AGB stars, with the abundance of HF being
essentially equal to the elemental abundance of fluorine. The
abundance of HF in the atmospheres of AGB stars, derived from
observations of ro-vibrational lines in the near-infrared, has
been used as a proxy of the abundance of fluorine (\cite{jor92}
1992). These authors found F abundance enhancements of up to a
factor of 30 over the solar value, and interpreted this as
evidence of $^{19}$F stellar nucleosynthesis in AGB stars. The
formation of fluorine remains poorly understood with three
different proposed sources: low-metallicity low-mass AGB stars,
Wolf-Rayet stars, and neutrino spallation in core-collapse
supernovae (see e.g. \cite{luc11} 2011). \cite{abi10} (2010) found
that the analysis of the HF ro-vibrational lines made by
\cite{jor92} (1992) was affected by line blends with C$_2$ and CN,
such that the derived F abundances are only slightly enhanced
(with a mean value of 0.2 dex) over the solar value, in closer
agreement with theoretical predictions of low-mass AGB stellar
models. In IRC +10216, we derive, through an alternative method
(the observation of the HF $J$ = 1-0 rotational line), a fluorine
abundance of 3.60 $\pm$ 0.30 (in the usual logarithmic scale with
H having a reference value of 12), which is almost one order of
magnitude lower than the solar value (4.56 $\pm$ 0.30;
\cite{asp09} 2009). This seems to indicate that in IRC +10216
there is an important degree of depletion of F onto dust grains,
even larger than that found in interstellar clouds where the HF
abundance relative to H$_2$ is $(1.1-1.6) \times 10^{-8}$
(\cite{son10} 2010; \cite{mon11} 2011). Our ability to place
constraints on the stellar nucleosynthesis of fluorine in IRC
+10216 is hampered by the HF abundance we derive being a lower
limit to the elemental abundance of fluorine. Nevertheless, the
low HF abundance derived here does not point toward any F
abundance enhancement over the solar value in IRC +10216, thus
arguing in favour of the results of \cite{abi10} (2010). Further
observations of the $J$ = 1-0 line of HF in other AGB stars will
help us to understand the degree of fluorine depletion onto dust
grains and any possible abundance enhancement due to stellar
nucleosynthesis.

In the case of hydrogen chloride, in the inner regions of IRC
+10216's envelope the derived abundance, including both isotopes,
is $1.1 \times 10^{-7}$ relative to H$_2$, making it the major
reservoir of chlorine, followed by AlCl whose abundance relative
to H$_2$ is $5 \times 10^{-8}$ (\cite{agu09} 2009). Other
chlorine-bearing species such as NaCl and KCl have lower
abundances, $1.3 \times 10^{-9}$ and $7 \times 10^{-10}$ relative
to H$_2$, respectively (\cite{agu09} 2009). Together, these
species have a total chlorine abundance of $8 \times 10^{-8}$
relative to the total number of H nuclei, i.e. one quarter of the
solar abundance of Cl ($3.2 \times 10^{-7}$ relative to H;
\cite{asp09} 2009). The abundance of chlorine should not be
modified inside AGB stars as $^{35}$Cl and $^{37}$Cl are thought
to be formed during explosive oxygen burning in supernovae
(\cite{woo73} 1973). Therefore, the missing chlorine in the inner
envelope of IRC +10216 may be in the form of neutral atomic Cl,
which together with HCl and AlCl is predicted to lock most of the
chlorine according to TE calculations (see Fig.~2 of \cite{cer10b}
2010b), or alternatively may have condensed onto dust grains.

\section{Concluding remarks}

The hydrogen halides HF and HCl have been proven to be ubiquitous
molecules in different types of interstellar environments. The
observations presented in this Letter show that these two
molecules are also important constituents of circumstellar
envelopes around evolved stars. They would be formed in the inner
regions close to the star under thermochemical equilibrium, which
predicts that HF and HCl should be the major reservoirs of
fluorine and chlorine, respectively, in the atmospheres of both O-
and C-rich AGB stars. The abundances derived here for the carbon
star envelope IRC +10216 are substantially lower than predicted by
TE calculations, suggesting that F and Cl are likely to be
severely depleted onto dust grains, by $\sim$90 \% for HF and
$\sim$75 \% for HCl, although some chlorine may also be in the
form of atomic Cl. We also find that in IRC +10216 the abundance
of F is probably not enhanced over the solar value by
nucleosynthesis in the AGB star, although this conclusion is
weakened somewhat by the HF abundance we derive being a lower
limit to the elemental abundance of fluorine. Both HF and HCl
should be detectable with HIFI through low $J$ pure rotational
transitions in other AGB stars, especially in those with high mass
loss rates, which have a large amount of emitting material and are
difficult to observe in the near-infrared owing to the opacity of
the dusty envelope.

\begin{acknowledgements}

HIFI has been designed and built by a consortium of institutes and
university departments from across Europe, Canada, and the United
States (NASA) under the leadership of SRON, Netherlands Institute
for Space Research, Groningen, The Netherlands, and with major
contributions from Germany, France and the US. Consortium members
are Canada: CSA, U. Waterloo; France: CESR, LAB, LERMA, IRAM;
Germany: KOSMA, MPIfR, MPS; Ireland: NUI Maynooth; Italy: ASI,
IFSI-INAF, Osservatorio Astrof\'isico di Arcetri-INAF;
Netherlands: SRON, TUD; Poland: CAMK, CBK; Spain: Observatorio
Astron\'omico Nacional (IGN), Centro de Astrobiolog\'ia
(INTA-CSIC); Sweden: Chalmers University of Technology - MC2, RSS
\& GARD, Onsala Space Observatory, Swedish National Space Board,
Stockholm University - Stockholm Observatory; Switzerland: ETH
Zurich, FHNW; USA: CalTech, JPL, NHSC. We thank G. Guillon and T.
Stoecklin for kindly providing the collision rate constants of HF
with H$_2$ prior to publication, and F. Lique and F. Dayou for
useful advice on inelastic collisions. We also acknowledge our
referee whose comments helped to improve this Letter. M.A., J. C.,
and F. D. thank the Spanish MICINN for funding support under
grants AYA2009-07304 and CSD-2009-00038. M.A. is supported by a
\textit{Marie Curie Intra-European Individual Fellowship} within
the European Community 7th Framework Programme under grant
agreement n$^{\circ}$ 235753.

\end{acknowledgements}

\end{document}